\documentclass[aps,prl,twocolumn,floatfix,superscriptaddress]{revtex4-1}
\usepackage{graphicx}
\usepackage{amsmath}
\usepackage{amssymb}
\usepackage{braket}
\usepackage{xcolor}
\usepackage{bm}
\usepackage[bookmarks=false,linkcolor=blue,urlcolor=blue,colorlinks,citecolor=blue]{hyperref}
\usepackage[normalem]{ulem}

\newcommand{\bea}{\begin{eqnarray}}
	\newcommand{\eea}{\end{eqnarray}}
\newcommand{\beq}{\begin{equation}}
	\newcommand{\eeq}{\end{equation}}

\begin{document}

\title{Majorana bound state parity exchanges in planar Josephson junctions}

\author{Varsha Subramanyan}
\affiliation{Department of Physics, University of Illinois at Urbana-Champaign, Urbana, IL USA}

\author{Jukka I. V\"{a}yrynen}
\affiliation{Department of Physics and Astronomy, Purdue University, West Lafayette, Indiana 47907, USA}

\author{Alex Levchenko}
\affiliation{Department of Physics, University of Wisconsin-Madison, Madison, Wisconsin 53706, USA}

\author{Smitha Vishveshwara}
\affiliation{Department of Physics, University of Illinois at Urbana-Champaign, Urbana, IL USA}

\begin{abstract}
We describe a scheme to exchange fermion parity between two pairs of Majorana bound states mediated by coupling with a centralized quantum dot. We formulate such a scheme for Majorana bound states nucleated in the Josephson vortices formed in a four-fold crossroads junction of planar topological superconductors in the presence of a perpendicular magnetic field. This platform yields several advantages to the execution of our scheme as compared to similar ideas proposed in wire geometries, including control over the positions of the MBS and hence, a tunable coupling with the quantum dot. We show that moving the MBS along the junctions through voltage pulses can facilitate parity exchange via a two-step process, with intermediate projective measurements of the quantum dot charge. Thus, we formulate a way to achieve single qubit operations for MBS  in extended Josephson junctions through projective measurements of quantum dot charge. We also discuss the physical viability of our scheme with a particular focus on changes in quantum dot energy levels as a measurable indicator of the success of the scheme.  
\end{abstract}
\maketitle

The heightened pursuit for material platforms to host robust qubits and scalable schemes for achieving error-resistant quantum computing has elicited a surge of experimental and theoretical advances in condensed matter physics. In the realm of topologically protected qubits, Majorana bound states (MBS) in topological superconducting setups continue offering promise as front-runners \cite{Sarma2015}. 
The original nanowire geometries, while facing setbacks, laid down the foundations for integrating the fields of condensed matter, material physics and quantum information science in designing viable platforms, and computational schemes~\cite{Lutchyn2010,Oreg2010,Alicea_2012,Aguado2017,PhysRevMaterials.2.044202,2019NatCo..10.5128Z,yu2021non,Pasquale}. Lately, extended junctions have become popular platforms to host MBS even at low magnetic field strengths \cite{FuKane,FuPotter, Williams,Ye, Flensberg}. Early proposals involve proximity-induced superconductivity on the surface of topological insulators in 2D and 3D with an externally applied magnetic field resulting in spatially separated MBS nucleating in Josephson vortices \cite{FuKane,FuPotter,FuKane2009,Lutchyn2010,Stern2019,Meyer2015,Grosfeld2011,Pientka,Fornieri2019,ren2019topological,Hegde2020, Flensberg,Paudel}.

While the manipulation of MBS cannot achieve a universal set of topologically protected quantum gates, the primary step in all these efforts is to construct effective gate operations by braiding the Majorana, followed by reliable readout mechanisms for the same. Schemes for performing gate operations involve two approaches: braiding via the physical motion of MBS and measurement-based braiding over stationary MBS~\cite{Alicea2011,Bonderson08b}. While these directions have been thoroughly explored in the nanowire context, to ensure progress, it is imperative that these schemes be innovatively adapted to a range of MBS settings~\cite{ zhou_fusion_2022, PhysRevLett.117.077002, PhysRevB.105.224509,Konakanchi2022}. 

Here we propose a scheme for MBS braiding and related fermion parity qubit operations that cater to the challenges and virtues of extended topological Josephson junction geometries. The scheme involves shuttling multiple MBS along such junctions towards a quantum dot that forms a parity measurement apparatus, thus forming a hybrid between the two gate operation  approaches~\cite{PRXQuantum.3.020340,PhysRevB.92.075143,PhysRevB.92.075143,PhysRevB.98.205403,PhysRevB.103.205427}. In what follows, we first outline the scheme. We next detail the physics of Josephson junction vortices as a natural pathway for nucleating MBS and efficiently manipulating them via the application of magnetic field pulses and local currents. We then discuss the workings of the quantum dot as relevant to our scheme. Finally, we bring these components together to present the specific manipulations and measurements that together form a controlled scheme for parity qubit operations.

\begin{figure}
    \centering
    \includegraphics[width=0.45\textwidth]{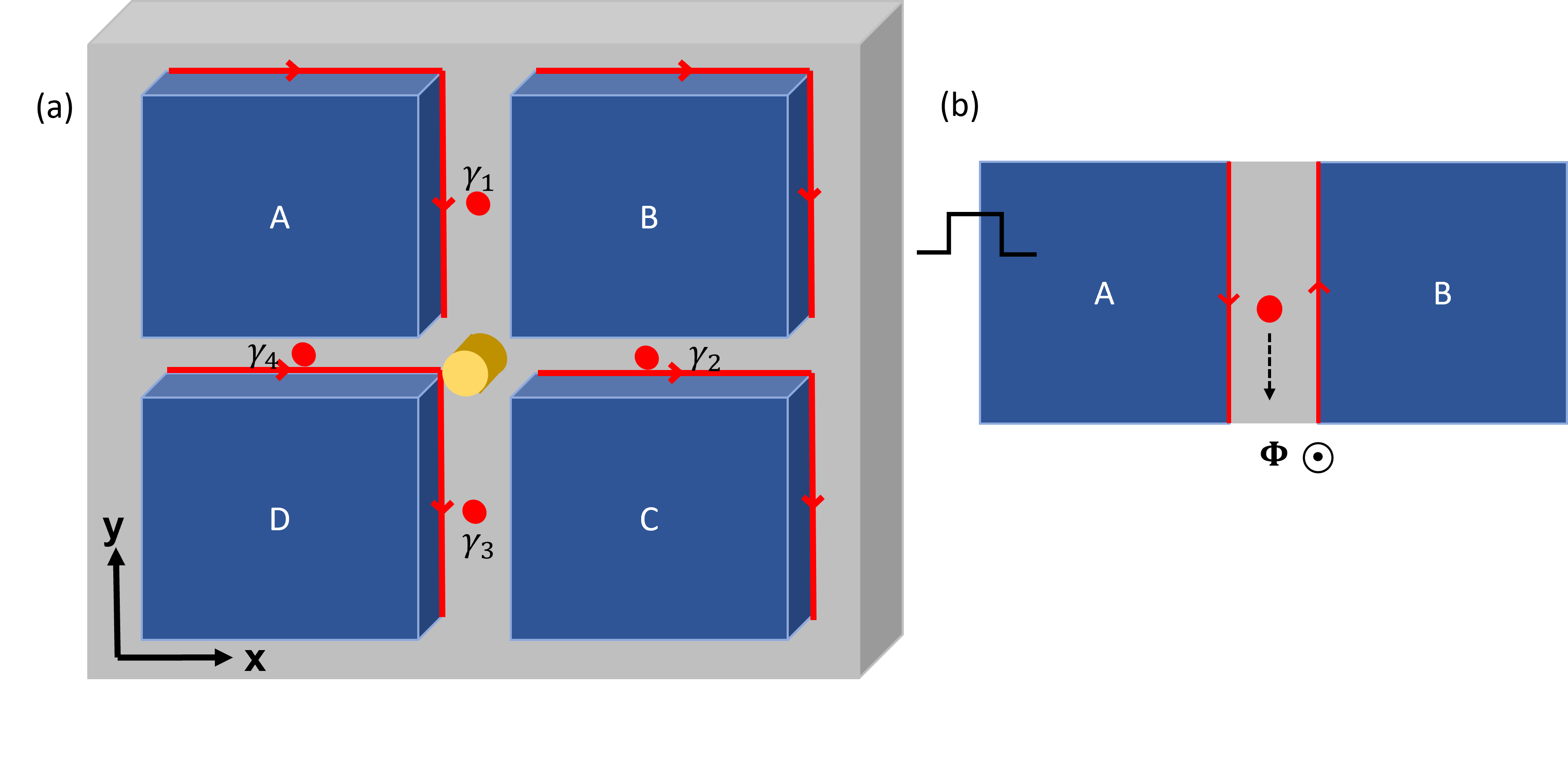}
    \caption{(a) A four-point crossroads junction of $p+ip$ superconducting islands separated by an insulator. Each junction is penetrated by sufficient magnetic flux so as to host at least one MBS (red dots). The central quantum dot (in yellow) couples to the MBS in each junction. The junction MBS is formed by hybridizing  chiral Majorana boundary states circulating the superconducting islands (red arrows).
    (b) Depiction of MBS motion within a particular junction when a biasing voltage pulse is applied. For magnetic flux $\Phi$ out of the page, and positive voltage $V$ across $A$ and $B$, the MBS will traverse in the negative $y$ direction.}
    \label{fig:islands}
\end{figure}

\textit{\textbf{ Outline of Scheme:}} The MBS-based scheme forming the core of this work is as follows. Our geometry consists of a four-point crossroads formed by four extended junctions between superconducting islands as shown in Fig. \ref{fig:islands}. Each of the four junctions has sufficient magnetic flux to nucleate at least one Majorana bound state (MBS). The center of the crossroads has a quantum dot that can be coupled to each of these 4 MBS. The extended junctions provide smooth channels for shuttling MBS via application of local voltage/current pulses across the bordering superconductors. The proximity of each MBS to the central dot controls the tunnel coupling between them. Coulomb blockade physics within the dot serves to dictate transfer of electrons between pairs of MBS and the dot as well as to act as a readout. 

For a pair of MBS $\gamma_i$ and $\gamma_j$, as commonly done, we can define the complex fermion operator $\Gamma_{ij}=\frac{1}{2}(\gamma_i-i\gamma_j)$ \cite{KITAEV}. The occupation number of this non-local electronic state $\hat{N}_{ij}=\Gamma^\dagger_{ij}\Gamma_{ij}$ has two eigenvalues (0 and 1) and defines the  parity $P_{ij} = 2 \hat{N}_{ij} - 1 = (-1)^{\hat{N}_{ij}}$ of that pair of MBS.  
Since we have 4 MBS in our system, the corresponding Hilbert space is 4-dimensional, and  the 3 possible ways of choosing the pairs of complex fermions correspond to different basis choices for the Hilbert space. In this work, parity exchanges take place between pairs of MBS mediated by the quantum dot, which is tuned to a regime of operation where it is treated as a two-level system that can be occupied or unoccupied. 

 Our scheme consists of two stages. Firstly, a pair of MBS with parity 1 (say $\gamma_i$ and $\gamma_j$) are coupled to the quantum dot, with the interaction resulting {\color{black} in a change in their parity through repeated measurements of the quantum dot occupation.} Secondly, a different pair of MBS (that is, at least one of the two MBS is different from $\gamma_i$ and $\gamma_j$) are coupled to the quantum dot, with this interaction resulting in changing their non-local parity to 1. 
As a specific example, we consider two canonical charge transfers - (a) Parity transfer from $\ket{1}_{12}\ket{0}_{34}$ to $\ket{0}_{12}\ket{1}_{34}$ and (b) Parity transfer from $\ket{1}_{12}\ket{0}_{34}$ to $\ket{0}_{13}\ket{1}_{24} = \frac{1}{\sqrt{2}}( \ket{0}_{12}\ket{1}_{34}  -i \ket{1}_{12}\ket{0}_{34} )$~\cite{Nayak,Ivanov}. All other combinations would be trivial variations or combinations of these two transformations. {\color{black} While the interaction of any two MBS with the quantum dot would be a non-destructive fusion process, the two-step process described here produces a net rotation in the 4-dimensional Hilbert space of qubits, as described by the Ivanov unitaries\cite{Ivanov}. Thus, non-Abelian rotation is achieved here. }Observation of such parity exchanges would correspond to landmark demonstrations of parity qubit manipulation and non-Abelian rotations in the degenerate qubit subspace.

\textit{\textbf{ MBS in the extended junctions:}} We model the extended junction geometry of Fig.~\ref{fig:islands} in the framework of the Bogoliubov-de Gennes (BdG) Hamiltonian for a spinless two-dimensional topological superconductor~\cite{Stone2004,Hegde2020,Abboud}, given by 
\begin{align}
	\mathcal H = \begin{pmatrix} -\frac{1}{2m^*}\bm \nabla^2 - \mu & \hat \Delta \\ \hat \Delta^\dag & \frac{1}{2m^*}\bm \nabla^2 + \mu  \end{pmatrix}. \label{hbdg}
\end{align}
The ${p_x+ip_y}$ chiral pairing operator is of the form ${\hat \Delta = i\frac{\Delta(\bm r)}{k_F} e^{i\varphi(\bm r)/2}(-i\partial_x+\partial_y)e^{i\varphi(\bm r)/2}}$, where  $\varphi(\bm r)$ denotes the superconducting phase. 

To first focus on a single extended junction, say, between islands $A$ and $B$, each island has a low-energy chiral Majorana state running along the boundaries of the superconductor (indicated in red in Fig. \ref{fig:islands}). The Josephson coupling between islands can be described by an effective Hamiltonian obtained by projecting onto the space spanned by their respective chiral dispersive Majorana edge modes \cite{FuKane, Grosfeld2011}. This effective Hamiltonian is given by 
\begin{align}
	\hat H_\text{edge} &= \frac{1}{2}\int ds\ \begin{pmatrix} \psi_A & \psi_B \end{pmatrix} \begin{pmatrix} iv \partial_s && -iW(s) \\ iW(s) && -iv \partial_s \end{pmatrix}
     \begin{pmatrix} \psi_A \\ \psi_B \end{pmatrix}.\label{heff}
\end{align}
Here  $\psi_{A}(s)$ and $\psi_B(s)$ are chiral Majorana fermions on the boundaries of island $A$ and 
$B$, and ${W(s) = \Theta(L-s)\Theta (s) m \cos\bigl(\frac{\phi(s)}{2}\bigr)}$ is the Josephson coupling term between the chiral Majoranas in the junction region $0<s<L$, of length $L$~\cite{Hegde2020,Abboud}. The superconducting phase difference between the islands is given by $\phi(s)=\varphi_R(s)-\varphi_L(s)$ and the variable $s$ is the coordinate along the boundaries of the islands. A small magnetic flux $\Phi_{AB}$ is now applied in the junction region. Since the Josephson coupling is only present in the region $0<s<L$, we will use the coordinate $y$ for junction $AB$ to focus on this region. 

In the short junction limit~\cite{tinkham2004} (when the length $L$ of the junction is shorter than the Josephson coherence length, or equivalently $J_c$ the Josephson current density is much lesser than  $\Phi_0/2\pi\mu_0L^2H$ for junctions of width $H$), the gauge-invariant superconducting phase difference is $\phi(y) = 2\pi\frac{\Phi}{\Phi_0}\frac{y}{L} +\phi_0$ 
defined up to a constant $\phi_0$. In general this constant could be determined by external phase bias (voltage/current/flux in a solenoid) between the two superconductors. Josephson vortices trapping MBS are obtained whenever $\phi(y)$ is an odd multiple of $\pi$ \cite{FuKane,FuPotter,Grosfeld2011,Hegde2020,Abboud}. Therefore, the number of MBS in each junction is given by the integer number of flux quanta in that junction. In this limit, the Josephson vortex solution takes the form of a large soliton~\cite{Grosfeld}. In the background of such soliton states, the zero energy eigenstates of Eq.~\eqref{heff} are MBS solutions of the form $(f[\phi(y)],g[\phi(y)])^T$. Here, the functions $f(y)$ and $g(y)$ are of the form $f(y)=A\sinh(\int^y\cos\phi(y')dy')-B\cosh(\int^y\cos\phi(y')dy')$ and $g(y)=A\cosh(\int^y\cos\phi(y')dy')-B\sinh(\int^y\cos\phi(y')dy')$, with specific constants determined by boundary conditions and external flux quanta (if any) penetrating the islands. 

In the presence of DC voltage $V$ applied across two islands \cite{FuPotter,Abboud} (say $A$ and $B$) for time $\Delta t$, the form of the phase difference across the junction is modified by the AC Josephson effect to include an additional term of $\frac{2\pi}{\Phi_0}Vt$. This phase difference across the junction gives rise to mobile solitonic Majorana states $(f[ky+\omega t],g[ky+\omega t])^T$, where $k=\frac{2\pi}{L}\frac{\Phi}{\Phi_0}$ and $\omega=\frac{2\pi}{\Phi_0}V$.  The velocity of the MBS is thus $ v_\phi=\frac{VL}{\Phi}$. The time dependent nature of the problem modifies the zero energy eigenstates of Eq.~\eqref{heff} to include a modified ``boosted" mass factor $m\rightarrow \frac{m}{\sqrt{1-(v_\phi/v)^2}}$. Crucially, we see that the application of voltage bias across the junction results in the traversal of the MBS along the junction. For a magnetic field pointing upwards from the plane of the islands, and positive voltage bias $V$, this setup corresponds to the vortices in the $AB$ junction moving in the $-\hat{y}$ direction (See Fig.~\ref{fig:islands}(b)). The reversal of any of these features - direction of magnetic field, sign of voltage, or application of voltage on $B$ instead of $A$ - would result in the MBS moving in the opposite direction. It is notable that very high voltages (equivalently, high velocities) can result in the destruction of the MBS through quasiparticle poisoning or hybridization effects. Therefore, it is necessary to pick appropriate voltages that ensure adiabatic transport of MBS to maintain topologically nontrivial ground states. 

In the crossroads setup, we assume all four islands to be initially at the same external voltage/common ground. Thus, applying voltage $V$ on island A affects both $\phi_{AB}$ as well as $\phi_{AD}.$ This voltage pulse results in MBS in the AB junction moving in the $-\hat{y}$ direction and those in the AD junction moving in the $-\hat{x}$ direction. {\color{black}We also assume that the phase difference at the center of the junction is not an odd multiple of $\pi$ so that no MBS is nucleated in the center of the croassroads junction. The system can be tuned out of such a parameter regime by the application of a small bias voltage on any of the islands.} We use the convention described here to exchange non-local MBS fermion parity. In each case, we start with total fermion parity being odd $-\gamma_1\gamma_2\gamma_3\gamma_4=-1$, with all four MBS being uncoupled to the dot. The dot itself is unoccupied.

\textit{\textbf{ MBS-Quantum dot interaction:}} Turning to the physics of the central quantum dot, we assume the standard associated single-particle energy level spacing and charging energy. Considering a small quantum dot with large level spacing, it is enough to account for only the level nearest to the Fermi energy and describe the dot by the single-level Hamiltonian $ H_{\text{QD}}=\Delta\varepsilon\hat{n}_d + \varepsilon_0$ where $\hat{n}_d=d^\dagger d$ is the quantum dot's occupation, $\varepsilon_0$ is the constant ground state energy and $\Delta\varepsilon$ sets the charging energy, tunable by the gate voltage. We assume here that the superconducting island is large enough to neglect its charging energy.

The total Hamiltonian of the system is thus given by $H_{\text{Tot}}=H_{\text{QD}}+H_{\text{Edge}}+H_{\text{T}}$, 
where the last term describes single-electron tunneling 
between the quantum dot and the four MBS closest to it. 
It is specified most generally as $H_{\text{T}}=\sum_{i=1}^4 \gamma_i(\lambda_i d^\dagger - \lambda_i^*d).$ The coupling constants $\lambda_i$ are in general complex, and exponentially suppressed with increasing distance between the MBS $\gamma_i$ and the quantum dot. 

As a simple first step, we explore the interaction between a pair of Majorana bound states $(1,2)$ and a quantum dot to show the oscillation and hence exchange of parity between the two. In the complex fermion notation $\Gamma_{12}$, the relevant part of the tunneling Hamiltonian takes the form 
\begin{align}
     H_{\text{T}}^{12}&= \Gamma_+(\lambda_1 d^\dagger - \lambda_1^*d)-i\Gamma_-(\lambda_2 d^\dagger - \lambda_2^*d)\nonumber \\
     &=\Lambda_e\Gamma_{12}^\dagger d^\dagger+\Lambda_e^* d \Gamma_{12} +\Lambda_o\Gamma_{12}^\dagger d + \Lambda_o^* d^\dagger \Gamma_{12}\label{HT}
\end{align}
where $\Gamma_\pm=\Gamma_{12}^\dagger\pm\Gamma_{12}$, and the redefined complex coupling constants $\Lambda_e = \lambda_1 -i \lambda_2$ and $\Lambda_o =  -\lambda_1^* +i \lambda_2^*$. This Hamiltonian acts on the Hilbert space spanned by $\{\ket{0}_{12}\ket{1}_D,\ket{1}_{12}\ket{0}_D,\ket{1}_{12}\ket{1}_D,\ket{0}_{12}\ket{0}_D\}$, where the first ket subscript $12$ refers to  electronic states with the specified occupation numbers formed by the Majorana degrees of freedom and the second ket with subscript $D$ refers to those of the dot. Generally, this Hamiltonian can have time-dependent coupling, especially as MBS are being moved. We will consider the case where they are brought close to the dot and held there. Since this Hamiltonian conserves total parity of the system, without loss of generality, we consider the temporal dynamics of the odd parity state $\ket{0}_{12}\ket{1}_D$. Time evolving this state, we obtain 
\begin{equation}
\begin{split}
    |\Psi(t)\rangle &=e^{iH^{12}_{\text{T}}t}\ket{0}_{12}\ket{1}_D\\
&=\cos(|\Lambda_o|t)\ket{0}_{12}\ket{1}_D+ie^{i\theta}\sin(|\Lambda_o|t)\ket{1}_{12}\ket{0}_D\label{TimeEvol}        
    \end{split}
\end{equation}
where $\theta=\textnormal{Arg}(\Lambda_ot)$. If the state is allowed to evolve up to time $t=\frac{\pi}{2|\Lambda_o|}$, the state is fully transformed to $\ket{1}_{12}\ket{0}_D$. Further evolution of the quantum state can be arrested by turning off the coupling constant through moving the MBS away from the quantum dot. Thus, we obtain a time scale for full parity transfer between the Majorana states and the quantum dot. But this protocol would require very precise control. The accumulated phases can also affect the desired outcome in the scheme, making the transformation not topologically protected. { \color{black} Writing down the evolution of the joint state under the total Hamiltonian would add factors depending on the charging energy to the time scale, and this argument would still hold.} We therefore now turn to projective measurements\cite{Bonderson08b,Karzig} to induce parity exchanges. 

A weak tunnel-coupling in Eq.~\eqref{HT} can be used to perform a (weak)  measurement of the Majorana parity $i\gamma_1 \gamma_2$~\cite{QPaper}.   To illustrate this, 
we turn to the perturbative effect of the tunneling in Eq.~\eqref{HT} to the quantum dot Hamiltonian and evaluate the parity-dependent shift of quantum dot energy levels. 
We label the ground and excited states of the quantum dot {\color{black}Hamiltonian $H_QD$ as $\varepsilon_0$ and $\varepsilon_1=\Delta\varepsilon+\varepsilon_0$. }

{\color{black}Treating the tunnel coupling terms $\Lambda_o$ and $\Lambda_e$ as small parameters, we now use perturbation theory to evaluate their effect on the energy levels $\varepsilon_0$ and $\varepsilon_1$ of the quantum dot.} The modified energy levels {\color{black} up to second order in perturbation strength } take the form
\begin{align}
    \varepsilon_0'&=\varepsilon_0-\frac{|\Lambda_o|^2(1-N_{12})+|\Lambda_e|^2 N_{12}}{\Delta\varepsilon}\\
    \varepsilon_1'&=\varepsilon_1+\frac{|\Lambda_e|^2(1-N_{12})+|\Lambda_o|^2N_{12}}{\Delta\varepsilon}
\end{align}
where $\Delta\varepsilon=\varepsilon_1-\varepsilon_0$ and $N_{12} = \langle \Gamma_{12}^\dagger\Gamma_{12} \rangle $ is the initial parity of the MBS. It is notable then, that the parity-dependent shift in energy levels scales as $\frac{|\Lambda_e|^2-|\Lambda_0|^2}{\varepsilon_1-\varepsilon_0}$, and thus, is only measurable when the even and odd sector couplings have different magnitudes~\cite{QPaper,Steiner}. {\color{black} That is, the spectrum of the  quantum dot is now parity dependent. Over repeated measurements, the probability distribution describing the occupation of the dot is obtained as a function of $\Delta\varepsilon$. This distribution is different for the two possible parity states of the coupled Majorana, thus offering a way to measure the parity of the MBS by measuring the quantum dot. This repeated measurement process also collapses the MBS pair coupled to it. } On moving the MBS away from the dot, the perturbative shift vanishes. Since the quantum dot measures MBS parity in a non-destructive way involving no hybridization, the {\color{black} process} may be performed multiple times by repeating the cycle of turning on interaction by moving the MBS pair close to the dot and measurement of the charge, and then undoing the measurement if the desired outcome is not obtained by making a different pair interact with the dot. 

\begin{figure*}
    \centering
    \includegraphics[width=\textwidth]{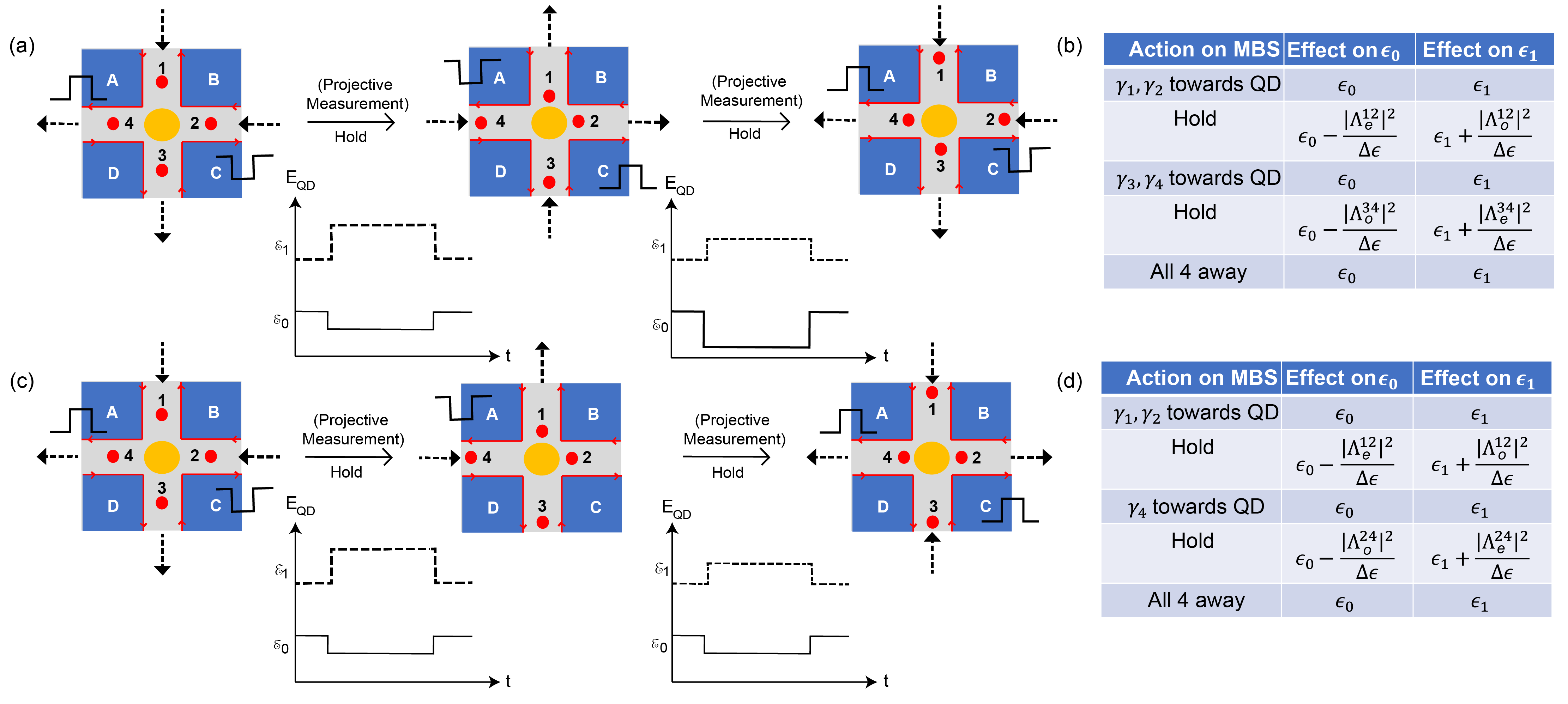}
    \caption{The motion of the MBS, the necessary voltage pulse sequences and the change in energy levels of the quantum dots are shown for each of the transformations: (a-b) $\ket{1}_{12}\ket{0}_{34}$ to $\ket{0}_{12}\ket{1}_{34}$ and (c-d) $\ket{1}_{12}\ket{0}_{34}$ to $\ket{0}_{13}\ket{1}_{24}$. In each case, a pair of MBS approach the quantum dot and are coupled to it. Through a projective measurement, the parity of the MBS pair is {\color{black} measured by} the quantum dot, as detected by changes in the quantum dot {\color{black} parity-dependent spectrum}. A different pair of MBS is then brought close to the quantum dot and a second projective measurement {\color{black} effectively transfers the parity to the second pair of MBS}, signalling the end of the scheme.} 
    \label{fig:scheme}
\end{figure*}

\textit{\textbf{ Qubit manipulation:}} To illustrate our key scheme, we can break up the steps corresponding to the transfer from $\ket{1}_{12}\ket{0}_{34}$ to $\ket{0}_{12}\ket{1}_{34}$ in the following way. {\color{black} We make the assumption that the total parity of the system is initially odd. Since the tunnelling Hamiltonian conserves parity, we expect the system to always remain in the odd parity sector. Let us say this initial state is $\ket{1}_{12}\ket{0}_D$}: (i) $\gamma_1$ and $\gamma_2$ are moved close to the quantum dot, and interact with it. {\color{black} The quantum dot measures the parity of the MBS as outlined in the previous section. The process is repeated (with undoing steps in between as necessary) until the dot-MBS system is in the desired state of $\ket{0}_{12}\ket{1}_D$.} (ii) The Majoranas $\gamma_1$ and $\gamma_2$ are moved away so that $\gamma_3$ and $\gamma_4$ can now move close to the quantum dot and interact with it. {\color{black} We repeat the measurement process to obtain the desired final state of  $\ket{1}_{34}\ket{0}_D$}. The voltage pulse sequence and corresponding QD spectrum are indicated in  Fig. \ref{fig:scheme}.

A similar break up is laid out for the transfer from $\ket{1}_{12}\ket{0}_{34}$ to $\ket{0}_{13}\ket{1}_{24}$. {\color{black} We assume the same initial state as before here}: (i) $\gamma_1$ and $\gamma_2$ are moved close to the quantum dot and interact with it. {\color{black} The quantum dot measures the parity of the MBS as outlined in the previous section. The process is repeated (with undoing steps in between as necessary) until the dot-MBS system is in the desired state of $\ket{0}_{12}\ket{1}_D$.} (ii) $\gamma_1$ is moved away and $\gamma_4$ is brought close to the quantum dot and interacts with it. {\color{black} We repeat the measurement process to obtain the desired final state of  $\ket{1}_{24}\ket{0}_D$}. The voltage pulse sequence and corresponding QD spectrum  are indicated in  Fig. \ref{fig:scheme}.

{\color{black} While the measurement of the parity of each individual pair of MBS resembles a (non-destructive) fusion process, the entirety of the scheme involves changing the state of the system from $\ket{1}_{12}\ket{0}_{34}$ to $\ket{0}_{12}\ket{1}_{34}$ or $\ket{1}_{12}\ket{0}_{34}$ to $\ket{0}_{13}\ket{1}_{24}$. That is, }these schemes perform net rotations over the degenerate subspace spanned by the states $\{\Gamma^\dagger_{12}\ket{00}, \Gamma^\dagger_{34}\ket{00}, \ket{00},\Gamma^\dagger_{12}\Gamma^\dagger_{34}\ket{00}\}$, with the first two states spanning the odd parity sector and the latter two spanning the even sector. Operations over this subspace may be represented in terms of unitary operators $U_{ij}=\frac{1}{\sqrt{2}}(1+\gamma_i\gamma_j)$ \cite{Ivanov}. The action of $U_{ij}$ corresponds to exchanging the MBS $\gamma_i$ and $\gamma_j$. The first transformation is equivalent to the action of the unitary $U=U_{21}U_{34}U_{23}U_{41}$ on the input state. Similarly, the second transformation is equivalent to the action of the unitary $U_{32}$ on the input state. We now analyze these projective measurements further, drawing an analogy with Stern-Gerlach experiments. Since the planar junction offers precise control over the coupling interaction through MBS motion, this scheme is better conceived in our geometry than in wire junctions where the MBS are stationary. 

When the quantum dot interacts with a pair of MBS (say $\gamma_1$ and $\gamma_2$), the quantum state exists in an entangled superposition of states with the same parity as shown in Eq.~\eqref{TimeEvol}. Our goal here, then, is to use the quantum dot to measure the non-local parity of the MBS pair, and in performing the measurement, effectively transfer parity from the MBS to the quantum dot \cite{Steiner}. It is useful to map the MBS parity operators to spin operators in the following way:
\begin{align}
    X=\frac{i}{2}\gamma_2\gamma_1,  \textnormal{ }Y=\frac{i}{2}\gamma_3\gamma_2,  \textnormal{ }Z=\frac{i}{2}\gamma_1\gamma_3. 
\end{align}
We can then rewrite the state of the system with arbitrary coefficients in terms of eigenstates of the $X$ operator-
\begin{align}
    |\psi \rangle=\alpha\ket{+}_X+\beta\ket{-}_X \,.
\end{align} 
{\color{black} We would like to project the state of the system such that the state of the system is $\ket{-}_X$. } {\color{black} It can be measured by observing the probability distribution of occupation of the quantum dot as described earlier. The probability of obtaining this outcome on measurement is given by $|\beta|^2$.}  If it is not obtained on measuring the dot, we "undo" the measurement by moving $\gamma_3$ close to the dot and $\gamma_2$ away from the dot by applying a positive voltage pulse on superconducting island $C$ {\color{black} so that this new pair of MBS interact with the quantum dot and are in superposition with it}. We then perform another charge measurement, which is equivalent to measuring the operator $Z$. {\color{black} Any outcome of this operation would yield a quantum state of the form $|\psi' \rangle=\frac{1}{\sqrt{2}}(\ket{+}_X\pm\ket{-}_X)$. } We can now revert to measuring $X$ by applying a negative voltage pulse on island $C$. {\color{black} During this second cycle of measuring the parity of X, the probability of the desired outcome is $\frac{1}{2}$. One can perform several cycles of such measurements until said desired outcome is obtained. }The probability of obtaining $\ket{-}_X$ for the first time on the $n$-th such cycle is $p_n(-)=|\beta|^2\delta_{n,1}+(1-\delta_{n,1})\frac{|\alpha|^2}{2^{n-1}}$. The total probability of obtaining $\ket{1}_D\ket{-}_X$ is thus calculated by adding up the probabilities for all $n$ as
\begin{align}
    p(-)=|\beta|^2+|\alpha|^2\sum_{n=2}^\infty\frac{1}{2^{n-1}}=|\beta|^2+|\alpha|^2=1
\end{align}
That is, after enough cycles, one is guaranteed to obtain the desired outcome. Thus, in each of the two schemes, replacing the hold times to projective measurements makes them less dependant on needing precise and time sensitive control for the scheme to work. 

Such measurements are performed over the crucial intermediate step involving the Rabi oscillation state given in Eq. \eqref{TimeEvol}. Were there infinitely precise control, one could move away the MBS precisely at time $t=\frac{\pi}{2|\Lambda_o|}$, smoothly changing the changing the state from $\ket{0}_{12}\ket{1}_{D}$ to $\ket{1}_{12}\ket{0}_{D}$, and preventing any further time evolution or phase accumulation.  However, in practice, the states accumulate various arbitrary dynamical phases over time evolution and interaction with the quantum dot, as seen in Eq. \eqref{TimeEvol}. Consider the second transformation for instance. Most generally, it can be written as $\ket{0}_{13}\ket{1}_{24}=U\ket{1}_{12}\ket{0}_{34}$ with $U=U_{32}U'(t)$ and $U'(t)=\textnormal{diag}(e^{i\alpha_1(t)},e^{i\alpha_2(t)},e^{i\beta_1(t)},e^{i\beta_2(t)})$ \cite{Ivanov,choi2019}. This would render these steps non topological, thus necessitating projective measurement in order to design protected gate operations. However, the possibility to accumulate such phases on specific qubit states will be useful in such geometries in designing a phase gate. Though not topologically protected, this will enable us in creating a universal set of quantum gates that can be realized on the extended junction platform..

\textit{\textbf{ Outlook:}} Turning to the experimental feasibility of our proposal, several components have been separately investigated and established. Josephson junction phase slips in extended geometries and response to fields and voltage pulses have a venerable history in a variety of superconductors. Applications of quantum dot physics in quantum information science is a vast area of study \cite{QD1,QD2,QD3,QD4,QD5,QD6,QD7}. More recently, extended topological junctions are receiving a concerted push with promising initial indicators of MBS physics \cite{Fornieri2019,ren2019topological,EJ0,EJ1,EJ3,EJ4,EJ5,EJ6}. 

The realization of our proposal would require a synergy of these disparate experimental elements working hand-in-hand with more involved theoretical treatments. In reality, even a single step, such as experimentally demonstrating adiabatic motion of vortices and associated MBS via STM measurements or coupling of MBS and the quantum dot via energy shifts and Rabi oscillation, would constitute a milestone. 

{\it Acknowledgements:} We are grateful to detailed discussions with Dale Van Harlingen, Guang Yue, Jessica Montone, Nick Abboud.
 We acknowledge the support of the National Science Foundation through Grant No. DMR-2004825 (VS and SV), Grant No. PHY-1748958 (VS), and the Quantum Leap Challenge Institute for Hybrid Quantum Architectures and Networks Grant No. OMA-2016136 (SV, AL). 
 This material is based upon work supported by the Office of the Under Secretary of Defense for Research and Engineering under award No. FA9550-22-1-0354 (JIV). 
 
\bibliography{ref}
\end{document}